\documentclass[pra,twocolumn,superscriptaddress]{revtex4}
\usepackage{amsmath}
\usepackage{amssymb}
\usepackage{color}
\usepackage[dvips]{graphicx}

\begin{document}

\title{Transverse Entanglement Migration in Hilbert Space}

\author{K.W. Chan}
\affiliation{The Institute of Optics, University of Rochester,
Rochester, NY 14627 USA}

\author{J.P. Torres}
\affiliation{ICFO-Institut de Ci\`{e}ncies Fot\'{o}niques, and
Department of Signal Theory and Communications, Universitat
Polit\`{e}cnica de Catalunya, Barcelona, Spain}

\author{J.H. Eberly}
\affiliation{Department of Physics and Astronomy, University of
Rochester, Rochester, NY 14627 USA}

\begin{abstract}
We show that, although the amount of mutual entanglement of
photons propagating in free space is fixed, the type of
correlations between the photons that determine the entanglement
can dramatically change during propagation. We show that this
amounts to a migration of entanglement in Hilbert space, rather
than real space. For the case of spontaneous parametric down
conversion, the migration of entanglement in
transverse coordinates takes place from modulus to phase of the
bi-photon state and back again. We propose an experiment to
observe this migration in Hilbert space and to determine the
full entanglement.
\end{abstract}

\maketitle

Entanglement is one of the truly central features of the quantum
world, and it forms the core of many applications based on quantum
theory. The observation of entanglement is generally achieved
through the measurement of correlations between entangled
subsystems. Correlation in quantum systems takes many forms and is
open to observation in a variety of ways. Therefore, the
determination of the amount of entanglement of quantum states
depends on the measurement of the correlations where entanglement
resides. This is of paramount importance, since in some experimental
configurations one registers types of correlation that might not be
appropriate to quantify the entangled nature of the quantum state.

In this Communication, we show that the measurement of correlation
between paired photons can miss the detection of entanglement
entirely. The underlying reason is an interesting
migration of entanglement that occurs in Hilbert space, but that
depends on coordinate location in real space. This is manifest in
photon correlations that show a rich and complex structure that
evolves during propagation, although the amount of entanglement is
constant. We focus here on entanglement that can become partly or
entirely identified with the phase of the state, in which case the
measurement of intensity correlations partially or completely
misses the existing entanglement. This is an observable
manifestation of the ``phase entanglement" previously
noted~\cite{chan} for massive particle breakup in an
Einstein-Podolsky-Rosen (EPR) scenario.

Entangled photons generated in spontaneous parametric
down-conversion (SPDC) are particularly open to the observation of
this phenomenon. The generated two-photon states have been shown
to exhibit entanglement in transverse momentum~\cite{burnham} and
in orbital angular momentum~\cite{mair,arnaut}. Moreover, one can
enlarge the Hilbert space of the two-photon state by using several
degrees of freedom~\cite{hyperentanglement}. The spatial
transverse degrees of freedom of photon pairs produced in SPDC
have attracted great attention because of the vast Hilbert space
involved~\cite{torres1,law2}, and the availability of techniques
to implement the $d$-dimensional quantum channel~\cite{torres2,howell1,zeilinger1}.

Observations of SPDC entanglement have usually been made either in
the near zone or the far zone~\cite{howell2}. Interestingly, in
the course of photon propagation from the near field zone to the
far field zone, the entanglement embedded in the two-photon
positional amplitude migrates out of the positional wave
function's modulus into its phase, and then back again.

In the region between near and far zones, the
entanglement not obtained through the measurement of intensity
correlations can be recovered by measuring the phase information
of the joint wave function.  Here we propose an experimental setup
to accomplish this by exploiting the symmetries of the wave
function.

We consider a nonlinear optical crystal of length $L$, illuminated
by a quasi-monochromatic laser pump beam, propagating in the $z$
direction. The signal and idler photons generated propagate from
the output face of the nonlinear crystal under the sole effect of
diffraction. The quantum state of the two-photon pair generated in
SPDC, at a distance $z$ from the output face of the nonlinear
crystal ($z=0$), reads in wave number space as
$|\Psi (z)\rangle = \int d\vec{p} \ d\vec{q} \ \ \Phi (\vec{p}, \vec{q},z) \
a_s^\dagger(\vec{p}) a_i^\dagger(\vec{q}) |0,0\rangle$, where
$\vec{p}$ and $\vec{q}$ are the transverse wave numbers of the
signal and idler photons, and $a_s^\dagger (\vec{p})$ and
$a_i^\dagger(\vec{q})$ are the corresponding creation operators.
The signal and idler photons are assumed to be monochromatic. This
assumption is justified by the use of narrow band interference
filters in front of the detectors.

Under conditions of collinear propagation of the pump,
signal and idler photons with no Poynting vector walk-off, which would
be the case of a noncritical type-II
quasi-phase matched configuration,
the mode function $\Phi(\vec{p}, \vec{q},z)$ is given by
\vspace{-1mm}
\begin{eqnarray}
\label{state1}
     \Phi(\vec{p}, \vec{q},z) & = & N E_p(\vec{p}+\vec{q})
     ~\text{sinc} \Big( \frac{\Delta_k L}{2} \Big)
      \exp \Big( i \frac{s_k L}{2} \Big) \nonumber \\
      & & \times \exp \{ i \left[ k_s
     (\vec{p}) + k_i
     (\vec{q}) \right]\,z \} ,
\end{eqnarray}
where $N$ is a normalization factor,
$\Delta_k = k_p(\vec{p} + \vec{q}) - k_s (\vec{p}) - k_i (\vec{q})$
and $s_k = k_p ( \vec{p} + \vec{q}) + k_s(\vec{p}) + k_i (\vec{q} )$,
$E_p$ is the transverse profile of the pump at the input face of the nonlinear
crystal, and $k_{j}$ ($j=p,s,i$) are wave number for the pump,
signal, and idler waves. We have also made use of the paraxial
approximation to describe the propagation of the signal and idler
photons in free space.

The sinc function that appears in Eq.~(\ref{state1}) can be
approximated by a Gaussian exponential function
which accurately retains the main features of the entanglement of the wave function~\cite{law2}.
We take $\text{sinc} \,bx^2 \simeq \exp [-\alpha b x^2]$
with $\alpha=0.455$, so that both functions coincide at the
$1/e^2$ intensity. Here we assume a pump beam with a Gaussian
shape. Therefore, the mode function can be written as~\cite{walborn, fedorov}
\begin{eqnarray}
\label{state2} \Phi(\vec{p}, \vec{q},z) \hspace{-1mm} &=&
\hspace{-1mm}
N \exp \hspace{-0.5mm}
\left\{\hspace{-0.5mm} -\frac{1}{4}\left[
\frac{w_0^2}{1+w_0^4/\sigma_0^2} + i \mu_1(z)
\right]|\vec{p}+\vec{q}|^2 \right\}
\nonumber \\
& &
\times
\exp \hspace{-0.5mm}\left\{
-\frac{1}{4}\left[\frac{\alpha L}{k_p^0}+ i \mu_2(z)\right] |\vec{p}-\vec{q}|^2
\right\} ,
\end{eqnarray}
where
$ N = \left\{(w_0^2\alpha L) / [\pi^2 (1+w_0^4/\sigma_0^2)
k_p^0] \right\}^{1/4}$,
$\mu_1(z) = 2(z + L)/k_p^0 - \sigma_0/(1+\sigma_0^2/w_0^4)$
and $\mu_2(z) = (2z + L)/k_p^0$. We denote $w_0$ as the pump beam
width and $\sigma_0 = -2R/k_p^0$, with $R$ being the radius of
curvature of the Gaussian beam at the entrance face of the
nonlinear crystal and $k_{p}^{0} = \omega_p n_p /c$. $\omega_p$
and $n_p$ are the corresponding angular frequency and refractive
index, respectively. Notice that we have made use of the
approximation $k_p^0 \simeq 2k_s^0 = 2k_i^0$. Moreover, all phase
factors independent of the transverse variables have been
neglected.

Equation~(\ref{state2}) describes the two-photon quantum state in
transverse wave number space $(\vec{p},\vec{q})$. We can also
describe the two-photon quantum state in coordinate space. In this
case, $\Psi(\vec{x}_s,\vec{x}_i, z) = 1/(2\pi)^2 \int d\vec{p}\
d\vec{q}\ \ \Phi(\vec{p}, \vec{q},z) \exp (i\vec{p} \cdot \vec{x}_s
+i\vec{q} \cdot \vec{x}_i)$, and since Eq.~(\ref{state2}) can be
written as $\Phi(\vec{p}, \vec{q}, z) = F(\vec{p}+\vec{q}, z)
G(\vec{p} - \vec{q}, z)$, one can easily obtain
\begin{eqnarray}
\label{central2}
    \Psi(\vec{x}_s,\vec{x}_i,z )
\hspace{-1mm} &=& \hspace{-1mm}
    N'
    \exp \left\{\hspace{-0.5mm} -\frac{1}{4\beta(z)} \hspace{-1mm}
    \left[ \frac{\alpha L}{k_p^0} - i \mu_2(z) \right] \hspace{-1mm}
    \left|\vec{x}_s-\vec{x}_i \right|^2 \hspace{-0.5mm} \right\}
\nonumber\\
&& \hspace{-17mm}
    \times
    \exp \left\{\hspace{-0.5mm} -\frac{1}{4\gamma(z)} \hspace{-1mm}
    \left[ \frac{w_0^2}{1+w_0^4/\sigma_0^2} - i\mu_1(z) \right] \hspace{-1mm}
    \left|\vec{x}_s+\vec{x}_i \right|^2 \hspace{-0.5mm} \right\} \hspace{-0.5mm}
,
\end{eqnarray}
where $N' = N / (\gamma \beta)^{1/4}$,
$\beta(z)=( \alpha L/k_p^0 )^2+ \mu_2^2(z)$, and
$\gamma(z)= w_0^4/(1+w_0^4/\sigma_0^2)^2 + \mu_1^2(z)$.
The conditional coincidence rate in
coordinate space is given by $\mathcal{R}_x(\vec{x}_s,\vec{x}_i,z)
= |\Psi(\vec{x}_s,\vec{x}_i,z)|^2$, while the conditional
coincidence rate in momentum space is
$\mathcal{R}_p(\vec{p},\vec{q},z) = |\Phi(\vec{p},\vec{q},z)|^2$.

Equation~(\ref{state2}) shows that the two-photon state is separable in
momentum space, and in coordinate space, if the conditions
$w_0^2/(1+w_0^4/\sigma_0^2) = \alpha L/k_p^0$ and
$\sigma_0/(1+\sigma_0^2/w_0^4) = L/k_p^0$ are fulfilled, which
corresponds to separability in modulus and phase.
Notwithstanding, from Eq.~(\ref{central2}) we also observe that it
is possible that the bi-photon function is separable in modulus
at a specific location, although not in phase. Therefore, $|\Psi
(\vec{x}_s,\vec{x}_i,z_0)|^2 = |\Psi_s (\vec{x}_s, z_0)|^2
|\Psi_i(\vec{x}_i,z)|^2$.

The central point is that, at a certain location $z_0$ from the
output face of the nonlinear crystal, where the $z$-dependent
condition $k_p^0 w_0^2 \beta (z_0) = (1 + w_0^4/\sigma_0^2)\alpha L \gamma(z_0)$ is
fulfilled, one does not observe intensity correlations at all in
coordinate space, although the quantum state is not separable in
either momentum or coordinate. Since the amount of entanglement
is determined by the existing correlations of the bi-photon
function in modulus and phase, at $z_0$ all entanglement lives
in the phase of the bi-photon function in coordinate space.

\begin{figure}[!t]
\centering\includegraphics[height=5.8cm]{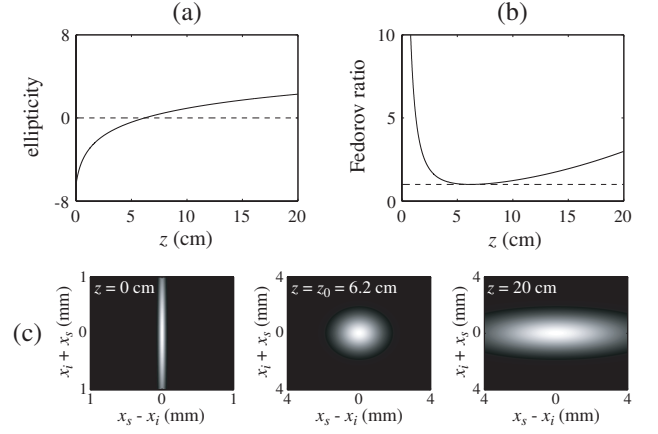}
\caption{ (a)
Ellipticity of the bi-photon function in coordinate space. (b)
Fedorov ratio $\mathcal{F}_x$ in coordinate space. The dashed line
corresponds to ${\cal R}_x=1$. (c) The spatial conditional
coincidence rate $\mathcal{R}_x(\vec{x}_s,\vec{x}_i,z)$ at different
locations: $z=0$\,cm, $z=z_0=6.2$\,cm and at $z=20$\,cm. Parameters:
Crystal length $L=5$\,mm; pump beam width $w_0=800\, \mu$m; pump
beam wavelength $\lambda_p=800$\,nm. The ellipticity is plotted in
logarithmic scale.} \label{figure1}
\end{figure}

Figure~\ref{figure1}(a) shows the evolution of the intensity
correlations as a function of the distance $z$. We plot the
ellipticity ($e$) in the plane ($\vec{x}_s+\vec{x}_i$,
$\vec{x}_s-\vec{x}_i$) of the bi-photon function given by
Eq.~(\ref{central2}), i.e., $ e=k_p^0 w_0^2 \beta/[(1+w_0^4/\sigma_0^2)\alpha L \gamma
]$. The spatial conditional coincidence rate
$\mathcal{R}_x(\vec{x}_s,\vec{x}_i,z)$ for three specific
locations $z$ are also shown in the figure.
Note that the amount of entanglement does not depend on the
location $z$, while the magnitude of the intensity correlations
evolves with $z$, as shown by the variation of ellipticity in Fig.~\ref{figure1}.
Therefore, although the amount of entanglement is unchanged
with $z$, the type of correlation that determines the entanglement
is different at every location $z$.

In order to quantify the amount of entanglement of the two-photon
state, we perform the Schmidt decomposition~\cite{grobe, ekert} of
the bi-photon function given by Eq.~(\ref{state2}) at the output
face of the nonlinear crystal. As shown in Appendix A, the amount
of entanglement denoted $K$ (i.e., the Schmidt number for continua~\cite{grobe})
is given by
\begin{equation}
\label{k}
K=\frac{\left[ \Re{(A+B)} \right]^2+\left[
\Im{(A-B)}\right]^2}{\left[
\Re{(A+B)}\right]^2-\left[ \Re{(A-B)} \right]^2} ,
\end{equation}
where $A=w_0^2/(1+w_0^4/\sigma_0^2)+i \mu_1$ and $B= \alpha L/ k_p^0+i\mu_2$.  Note
that $K$ does not depend on $z$ even though $\mu_1$ and $\mu_2$ do
so.

For the sake of comparison, let us consider the Fedorov
ratio~\cite{fedorov2}, here denoted $\mathcal{F}$, a typical
correlation measurement that could potentially be employed to show
the existence of entanglement. For the signal photon in momentum
space it takes the form
$\mathcal{F}_{s,p} \equiv \langle\Delta^2 \vec{p_s} \rangle / \langle
\Delta^2 \vec{p_s}\rangle_{i}$, and the expression in coordinate
space is $\mathcal{F}_{s,x} \equiv \langle \Delta^2 \vec{x}_s \rangle
/ \langle \Delta^2 \vec{x}_s \rangle_{i}$. Here the variance
averages not containing subscript $i$ are unconditional. The
averages with subscript $i$ are conditioned on the idler photon,
which is to be constrained by $\vec p_i = 0$ and  $\vec x_i = 0$,
in the $\vec p_s$ and $\vec x_s$ averages respectively.

If the entanglement resides only in the modulus of the bi-photon given by Eq.~(\ref{state2}), i.e., $\mu_1=\mu_2$,
$\mathcal{F}_p$ can be shown to be equal to the amount of entanglement given by Eq.~(\ref{k}), while $\mathcal{F}_x$ only gives
the correct amount of entanglement in the near and far fields. This is the typical experimental condition if the pump beam
shows no curvature at the input face of the nonlinear crystal.

If part or all the entanglement resides in the phase of the
bi-photon, even $\mathcal{F}_p$ does not correctly measure the amount
of the entanglement of the quantum state, only the part of the
entanglement that resides in the modulus of the bi-photon
function. Figure~\ref{figure1}(b) shows the Fedorov ratio in coordinate space
for a typical case. At $z = z_0$, where the bi-photon function
shows no ellipticity in the modulus, we have $\mathcal{F}_x = 1$,
although the quantum state is entangled.
\begin{figure}[!t]
\centering\includegraphics[width=7cm]{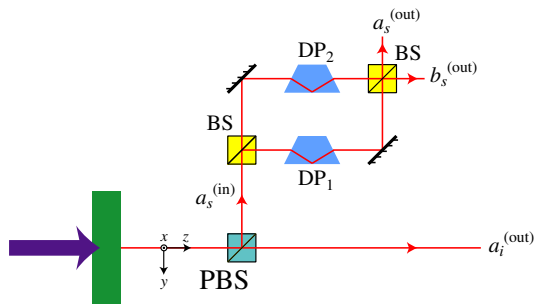}
\caption{(Color online)
Experimental scheme to detect the total entanglement.
The signal photon is directed into a
modified Mach-Zehnder interferometer with
two Dove prisms $\text{DP}_i$ with orientation angles $\theta_i$.
PBS is a polarization beam splitter.}
\label{figure2}
\end{figure}

In Fig.~\ref{figure2} we show an experimental scheme to detect the total
entanglement of the bi-photon described by Eqs.~(\ref{state2})
or~(\ref{central2}). The signal photon is sent to
a modified Mach-Zehnder interferometer with two Dove prisms
inserted in the interfering arms.  The arms are assumed to be
balanced so that the relative phase shift between the two arms of
the interferometer due to propagation is zero.

We set the orientation angles of the Dove prisms $\theta_1=
\pi/2$ and $\theta_2=0$. The conditional coincidence rates
of the output ports of the interferometer shown in Fig.~\ref{figure2}
take the form (see Appendix B)
\begin{subequations}
\label{mach1}
\begin{eqnarray}
& & P_{+}
= \iint d \vec{x}_s \, d\vec{x}_i \ P_{a_s,a_i} (\vec{x}_s, \vec{x}_i )
= \frac{1}{2} \left( 1+\frac{1}{K} \right) , \qquad \\
& & P_{-}
= \iint d \vec{x}_s \, d\vec{x}_i \ P_{b_s,a_i} ( \vec{x}_s, \vec{x}_i )
= \frac{1}{2} \left( 1-\frac{1}{K} \right) , \qquad
\end{eqnarray}
\end{subequations}
where
\vspace{-1mm}
\begin{subequations}
\label{mach2}
\begin{eqnarray}
&& \hspace{-4mm}
     P_{a_s,a_i}(\vec{x}_s, \vec{x}_i)
=     \frac{1}{4}
     \left|\Psi(x_s, y_s; \vec{x}_i)
     + \Psi(-x_s, -y_s; \vec{x}_i)\right|^2 \hspace{-1mm}, \hspace{5.5mm}
\\
&& \hspace{-4mm}
     P_{b_s,a_i}(\vec{x}_s, \vec{x}_i)
=      \frac{1}{4}
     \left|\Psi(x_s, -y_s; \vec{x}_i)
     - \Psi(-x_s, y_s; \vec{x}_i)\right|^2 \hspace{-1mm}. \hspace{5.5mm}
\vspace{-1.5mm}
\end{eqnarray}
\end{subequations}
Therefore, the amount of entanglement of the quantum
state given by Eq.~(\ref{state2}) can be quantified as $K = (P_{+}
+ P_{-})/(P_{+} - P_{-})$. The experimental setup plotted in
Fig.~\ref{figure2} measures the ``full'' entanglement of the quantum state
described by Eqs.~(\ref{state2}) or~(\ref{central2}).

\begin{figure}[!tb]
\centering\includegraphics[width=4.4cm]{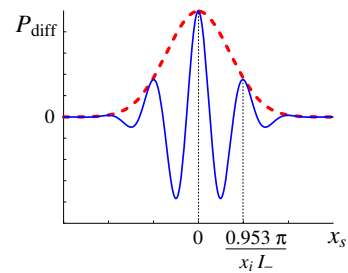}
\caption{(Color online)
The coincidence interference pattern observed in the setup of Fig.~\ref{figure2}.
The dotted line shows the situation when phase entanglement is absent.
\vspace{-3mm}}
\label{figure3}
\end{figure}

On the other hand, the joint probability distributions in
Eq.~(\ref{mach2}) also exhibit interesting interference behavior.
Using Eq.~(\ref{central2}), we have
\vspace{-1mm}
\begin{eqnarray}
     P_\text{diff}(\vec{x}_s, \vec{x}_i)
\hspace{-1mm} &\equiv& \hspace{-1mm}
     P_{a_s,a_i}(\vec{x}_s, \vec{x}_i)
     - P_{b_s,a_i}(\vec{x}_s, \vec{x}_i)
\nonumber\\
\hspace{-1mm} &=& \hspace{-1mm}
     \mathcal{N}^2 e^{-R_+(z) (\vec{x}_s^2 + \vec{x}_i^2)}
     \cos\left[2 I_-(z) \vec{x}_s\cdot\vec{x}_i\right] , \ \ \
\vspace{-1mm}
\label{eq:Paa+-Pba}
\end{eqnarray}
where $R_+(z) = w_0^2/\left[(1+w_0^4/\sigma_0^2)\gamma(z)\right] + \alpha L/\left[k_p^0
\beta(z)\right]$ and $I_-(z) = \mu_1(z)/\gamma(z) -
\mu_2(z)/\beta(z)$. The difference of the joint probability
$P_\text{diff}$ is plotted in Fig.~\ref{figure3} as a function of
$x_s$. It is seen that the location of the second maximum is at
$x_s \approx 0.953 \pi/x_iI_-$, from which we can determine the
entanglement in the phase.
That is, the interference of the wave function with itself,
with its symmetry, reveals not only the phase information of the wave
function, but more importantly how the two photons are correlated in the phase
through the fringe pattern given by Eq.~(\ref{eq:Paa+-Pba}).

In conclusion, we have demonstrated the Hilbert space migration of
entanglement of down-converted photons in free-space propagation.
We suggested its implication for experiments involving the
quantification of the degree of entanglement by means of common
variance measurements. We have also suggested a simple
experimental scheme that can detect both the entanglement in
amplitude and phase.

KWC acknowledges support from a fellowship from the Croucher
Foundation; JPT from the Generalitat de Catalunya, from the
European Commission under the integrated project Qubit
Applications (QAP) funded by the IST directorate (Contract
No.~015848), and from the MEC (Consolider Ingenio 2010 project {\em
QIOT} CSD2006-00019, FIS2004-03556); and JHE from ARO Grant
W911NF-05-1-0543. Discussions with R.W. Boyd, C.K. Law, M.N.
O'Sullivan-Hale and K. Wodkiewicz have been very helpful.

\noindent\emph{Appendix A} --- Let us consider a bi-photon
amplitude that, at the output face of the nonlinear crystal
($z=0$), is written as
\vspace{-1mm}
\begin{equation}
\label{newbiphoton}
\Phi(\vec p,\vec q) =
\Big[\frac{\Re(A)\Re(B)}{\pi^2}\Big]^{\frac{1}{4}}
\exp \Big(-\frac{A|\vec{p}_+|^2+B|\vec{p}_-|^2}{4} \Big) ,
\end{equation}
where $\vec{p}_+ = \vec{p}+\vec{q}$ and $\vec{p}_- =
\vec{p}-\vec{q}$. Inspection
of Eq.~(\ref{newbiphoton}) shows that
the bi-photon function can be separated for the two transverse
dimensions, i.e., $\Phi(\vec{p}, \vec{q}) = \Phi_x (p_x, q_x)
\Phi_y (p_y, q_y)$. Therefore, the Schmidt decomposition of
Eq.~(\ref{newbiphoton}) can be written as $\Phi (\vec{p}, \vec{q}) =
\sum_{m,n=0}^{\infty} \sqrt{\lambda_{mn}} f_{mn} ( \vec{p} )
g_{mn} ( \vec{q} )$, where the basis functions of the
decomposition are $f_{mn} (\vec{p}) = \psi_m (p_x) \psi_n (p_y)$
and $g_{mn} (\vec{q}) = \psi_m (q_x) \psi_n (q_y)$, with
eigenvalue $\sqrt{\lambda_m \lambda_n}$.

The reduced density matrix for the signal photon, $\rho_s
(p_x,p_y,\bar{p}_x,\bar{p}_y,) = \text{Tr}_\text{idler}\,|\Phi\rangle\langle\Phi|$, can be
separated into two matrices, i.e.,
$\rho_s(p_x,p_y,\bar{p}_x,\bar{p}_y,) = \rho_x ( p_x,\bar{p}_x)
\rho_y (p_y,\bar{p}_y,)$, each one corresponding to a transverse
coordinate. The functions $\psi_m$, and the eigenvalues
$\lambda_m$ can be obtained from the expansion of the one
dimensional reduced density matrix $\rho_x$ in the form $\rho_x
(p_x,\bar{p}_x ) = \sum_{n=0}^{\infty} \lambda_{x,n} \psi_n ( p_x )
\psi_n (\bar{p}_x)$.

From Eq.~(\ref{newbiphoton}), the one-dimensional reduced density
matrix becomes
\vspace{-2mm}
\begin{eqnarray}
\label{reduced}
\rho_x ( p_x,\bar{p}_x )
\hspace{-1mm} &=& \hspace{-1mm}
\sqrt{\frac{2 \Re{(A) \Re{(B)}}}{\pi \Re{(A+B)}}}
\exp \left\{-i\frac{\Im{(AB)} (p_x^2-\bar{p}_x^{2})}{2
\Re{(A+B)}} \right\}
\nonumber \\
& & \times \exp \left\{-[( a+b ) (p_x^2+\bar{p}_x^2 )-2b p_x
\bar{p}_x ] \right\} , \quad
\end{eqnarray}
where $a=\Re{(A)} \Re{(B)}/\Re{(A+B)}$ and $b = |A-B|^2/
[8\Re{(A+B)}]$. The representation given by Eq.~(\ref{reduced}) is
also found in the determination of the mode structure of
Gaussian-Schell model sources in the theory of partial
coherence~\cite{starikov}. One can find that the eigenvalue
$\lambda_{x,n}$ is given by $\lambda_{x,n} = (a/c)^{1/2}
(1-w^2)^{1/2} w^n$, with $c = (a^2+2ab)^{1/2}$ and $w =
b/(a+b+c)$. The functions $\psi_n$ are appropriately normalized
Hermite-Gaussian functions. Thereafter, the calculation of
$K=1/\sum_{m,n=0}^{\infty} (\lambda_{x,m} \lambda_{y,n})^2$ would
yield $K = (c/a)^2$, which can be straightforwardly written as the
expression that appears in Eq.~(\ref{k}). Notice that $K$
describes entanglement in the $(p_x,p_y)$ space.

\vspace{2mm}

\noindent\emph{Appendix B} --- The modified Mach-Zehnder
interferometer shown in Fig.~\ref{figure2} contains three basic
elements that modify the spatial shape of the bi-photon function.
The action of the mirrors is described by $\hat{a}_{in} ( x,y )
\rightarrow \hat{a}_{out} ( x, - y )$, where $x$ and $y$ are the
transverse coordinates in the frame of \emph{each} individual
beam.

The action of the beam-splitter is~\cite{walborn}
$\hat{a}_{in}(x,y) \rightarrow \hat{a}_t (x,y) + i \hat{a}_r
(x,-y)$, where $\hat{a}_t$ is the creation operator of the
transmitted photon, and  $\hat{a}_r$ the corresponding creation
operator of the reflected photon. For a Dove prism that is rotated
by an angle $\theta$ with respect to the axis of image inversion,
the fields before and after the dove prism are given by
$\hat{a}_{in} ( x,y ) \rightarrow \hat{a}_{out} (x \cos 2\theta -
y\sin 2\theta, -x \sin2\theta - y \cos2\theta)$.
Together with the effect of the polarization beam-splitter, one thus obtains
that all joint probability detections in the configuration
described in Fig.~\ref{figure2} are given by Eq.~(\ref{mach2}).

The bi-photon function at location $z$ can be written in the form
\vspace{-3mm}
\begin{eqnarray}
\label{final1}
\Psi ( \vec{r}_s, \vec{r}_i )
\hspace{-1mm} &=& \hspace{-1mm}
\left[ \frac{a (1-w^2)}{c} \right]^{1/2}
\sum_{n,m=0}^{\infty} w^{n/2} w^{m/2} \nonumber \\[-0.5mm]
& & \hspace{-10mm}
\times \psi_m ( x_s,z ) \psi_n ( y_s,z ) \psi_m ( x_i,z ) \psi_n (y_i,z ) , \qquad
\vspace{-3mm}
\end{eqnarray}
where the function $\psi_n (x,z)$ corresponds to Hermite-Gaussian
function at $z=0$ that evolves under the sole influence of
diffraction. Due to the symmetry of the Hermite-Gaussian
functions, one has  $\psi_n ( -x,z) = (-1)^n \psi_n ( x,z )$.
Making use of this symmetry property,  one obtains
\vspace{-2mm}
\begin{equation}
\label{final2}
P_+ = \frac{a ( 1-w^2)}{2c}
\left[\sum_{n,m=0}^{\infty} w^{n+m} + (-w )^{n+m}\right] .
\vspace{-2mm}
\end{equation}
From Eq. (\ref{final2}), one obtains Eq. (\ref{mach1}), taking
into account that the amount of entanglement is given by $K =
(c/a)^2$.


\begin{thebibliography}{99}

\bibitem{chan} K.W. Chan and J.H. Eberly, arXiv:quant-ph/0404093.

\bibitem{burnham} D.C. Burnham and D.L. Weinberg, Phys. Rev. Lett.
\textbf{25}, 84 (1970).

\bibitem{mair} A. Mair, A. Vaziri, G. Weihs and A. Zeilinger, Nature \textbf{412}, 313 (2001).

\bibitem{arnaut} H.H. Arnaut and G.A. Barbosa, Phys. Rev. Lett. \textbf{85}, 286 (2001).

\bibitem{hyperentanglement} J.T. Barreiro, N.K. Langford, N.A. Peters,
and P.G. Kwiat, Phys. Rev. Lett. \textbf{95}, 260501 (2005).

\bibitem{torres1} J.P. Torres, A. Alexandrescu, and L. Torner, Phys. Rev. A \textbf{68}, 050301(R) (2003).

\bibitem{law2} C.K. Law and J.H. Eberly, Phys. Rev. Lett. \textbf{92}, 127903 (2004).

\bibitem{torres2} J.P. Torres, Y. Deyanova, L. Torner, and G. Molina-Terriza,
Phys. Rev. A \textbf{67}, 052313 (2003).

\bibitem{howell1} M.N. O'Sullivan-Hale, I. Ali Khan, R.W. Boyd,
and J.C. Howell, Phys. Rev. Lett. \textbf{94}, 220501 (2005).

\bibitem{zeilinger1} S. Groblacher, T. Jennewein, A. Vaziri, G. Weihs,
and A. Zeilinger, New J. Phys. \textbf{8}, 75 (2006).

\bibitem{howell2} J.C. Howell, R.S. Bennink, S.J. Bentley,
and R.W. Boyd, Phys. Rev. Lett. \textbf{92}, 210403 (2004).

\bibitem{walborn} S.P. Walborn, A.N. de Oliveira, S.
Padua, and C.H. Monken, Phys. Rev. Lett \textbf{90}, 143601 (2003).

\bibitem{fedorov} M.V. Fedorov, M.A. Efremov, P.A. Volkov,
E.V. Moreva, S.S. Straupe, and S.P. Kulik, arXiv:quant-ph/0612104.

\bibitem{grobe} R. Grobe, K. Rzazewski and J.H. Eberly, J. Phys. B
\textbf{27}, L503 (1994).

\bibitem{ekert} A. Ekert and P.L. Knight, Am. J. Phys. \textbf{63}, 415 (1995).

\bibitem{fedorov2} M.V. Fedorov, M.A. Efremov, A.E. Kazakov, K.W. Chan,
C.K. Law, and J.H. Eberly, Phys. Rev. A \textbf{72}, 032110 (2005).

\bibitem{starikov} A. Starikov and E. Wolf, J. Opt. Soc. Am. \textbf{72}, 923 (1982).

\end{thebibliography}
\end{document}